\documentclass[a4paper]{JAC2003}
\usepackage{graphicx}
\usepackage{booktabs}
\usepackage{varwidth}
\usepackage{xcolor}

\begin{document}
\title{BEAM--BEAM OBSERVATIONS IN THE RHIC\thanks{This work was supported by Brookhaven Science Associates, LLC under Contract No. DE-AC02-98CH10886 with the US Department of Energy.}}
\author{Y. Luo, W. Fischer, BNL, Upton, NY, USA }
\maketitle

\begin{abstract}
The Relativistic Heavy Ion Collider (RHIC) at Brookhaven National Laboratory has been in operation since 2000.  Over the past decade, the luminosity in the polarized proton (p-p) operations has increased by more than one order of magnitude. The maximum total beam--beam tune shift with two collisions has reached 0.018. The beam--beam interaction leads to large tune spread, emittance growth, and short beam and luminosity lifetimes. In this article, we review the beam--beam observations during the previous RHIC p-p runs. The mechanism for particle loss is presented. The intra-beam scattering (IBS) contributions to emittance and bunch length growths are calculated and compared with the measurements. Finally, we will discuss current limits in the RHIC p-p operations and their solutions.

\end{abstract}

\section{Introduction}

RHIC consists of two superconducting rings, the Blue ring and the Yellow ring. They intersect at six locations around the ring circumference. The beam in the Blue ring circulates clockwise and the beam in the Yellow ring circulates counterclockwise. The two beams collide at two interaction points (IPs), IPI6 and IP8.  Figure~1 shows the layout of the RHIC. The RHIC is capable of colliding heavy ions and polarized protons (p-p). The maximum achieved total beam--beam parameter with two collisions was 0.003 in the 100~GeV Au--Au collision and 0.018 in the p-p collision. In this article, we only discuss the beam--beam effects in the p-p runs.

The working point in the RHIC p-p runs is chosen to provide a good beam--beam lifetime and to maintain the proton polarization during the energy ramp and physics store. The current working point is constrained between 2/3 and 7/10:  2/3 is  a strong third-order betatron resonance; 7/10 is a 10th-order betatron resonance and also a spin depolarization resonance~\cite{Mei1}. Experiments and simulations have shown that the beam lifetime and the proton polarization are reduced when the vertical tune of the proton beam is close to  $7/10$.

The main limits to the beam lifetime in the RHIC p-p runs are the beam--beam interaction, the non-linear magnetic field errors in the interaction regions (IRs), the non-linear chromaticities with low $\beta^{\ast}$s, the horizontal and vertical third-order betatron resonances, and the machine and beam parameter modulations.

To further increase the luminosity, we can either increase the bunch intensity or reduce $\beta^*$.  Figure~2 shows the proton tune footprints including beam--beam interactions. In this calculation, the proton bunch intensity is $2.0\times10^{11}$ and the 95\% normalized transverse emittance is 15$\pi$~mm$\cdot$mrad.  The total beam--beam parameter with two collisions is 0.02. From Fig.~2, there is not enough tune space to hold the large beam--beam tune spread when the proton bunch intensity is larger than $2.0\times10^{11}$.

To minimize the beam--beam tune spread and to compensate the non-linear beam--beam resonance driving terms, head-on beam--beam compensation with electron lenses (e-lenses) is adopted for the RHIC~\cite{RHIC-elens1,OPPS,WF}. Two e-lenses are being installed on either side of IP10, one for the Blue ring and one for the Yellow ring. The goal of head-on beam--beam compensation is to double the current RHIC luminosity in the p-p operations.

\begin{figure}
\begin{center}
\includegraphics[width=55mm]{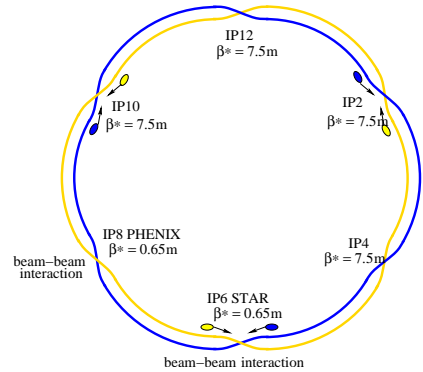}
\end{center}
\caption{The layout of the RHIC. Two beams collide at IP6 and IP8.}
\end{figure}

\begin{figure}
\begin{center}
\includegraphics[width=70mm]{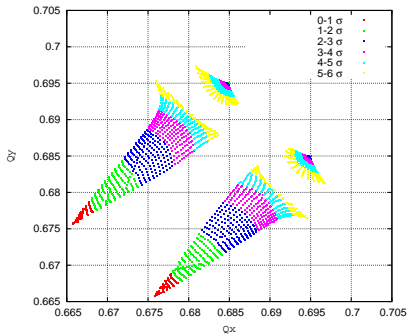}
\end{center}
\caption{Tune footprints without and with  beam--beam. The bunch intensity is $2.0\times10^{11}$. }
\end{figure}

\section{OBSERVATIONS}

\subsection{Previous p-p Runs}

The luminosity in the p-p collision is given by
\begin{equation}
L= \frac{3 N_{\rm p}^2 N_{\rm b} \gamma f_{\rm rev} }{2 \pi \epsilon_{\rm n} \beta^*}H(\frac{\beta^*}{\sigma_{\rm l}}).
\end{equation}
Here, $N_{\rm p}$ is the proton bunch intensity, $N_{\rm b}$  is the number of bunches,  $\gamma$  is the Lorentz factor, and $f_{\rm rev}$  the revolution frequency. $\epsilon_{\rm n}$ is the 95\% normalized emittance and $\sigma_{\rm l}$  the  r.m.s.\@ bunch length. $H(\frac{\beta^*}{\sigma_{\rm l}})$ is the luminosity reduction due to the hourglass effect. The total beam--beam parameter, or the total linear incoherent beam--beam tune shift with two collisions, is
\begin{equation}
\xi=\frac{3 N_{\rm p} r_{\rm p}}{ \pi \epsilon_{\rm n}}.
\end{equation}
Here, $r_{\rm p}$ is the classical radius of a proton. We have assumed two collisions, at IP6 and IP8.

In the 2009 RHIC 100~GeV p-p run, with $\beta^*$=0.7~m and a bunch intensity of $1.5\times10^{11}$, we observed a shorter beam lifetime of 7~h compared to 12 h in the 2008 RHIC 100~GeV p-p run with $\beta^*=1.0$~m~\cite{RHIC2005,RHIC2006,RHIC2008, RHIC2009}. In the 2012 RHIC 100~GeV p-p run, $\beta^*=0.85$~m lattices were adopted, and the beam lifetime was 16~h with a typical bunch intensity of $1.65\times10^{11}$~\cite{RHIC2012}.

In the 2011 250~GeV and 2012 255~GeV p-p runs, a common 9~MHz RF system was used to produce a long bunch length on the energy acceleration to maintain both transverse and longitudinal emittances~\cite{RHIC2012,RHIC2011}. When the beams reached store energy, the bunches were re-bucketed to the 28~MHz RF system. To  achieve an even shorter bunch length, we added 300~kV 197~MHz RF voltages  at store.  In these two runs, $\beta^*$ at the collision points was 0.65~m. The maximum bunch intensity reached $1.7\times10^{11}$. The store length was 8~h.  Table~1 shows the lattice and typical beam parameters in the 2012 RHIC 255~GeV p-p runs.

\begin{table}
\centering
\caption{Parameters in 2012 255~GeV p-p Runs}
\label{tab:rhicpar}
\begin{tabular}{lcc}
\hline\hline
Parameter                                          &  Unit       &    Value  \\  \hline\hline
No. of colliding bunches                            &   -          &     107     \\  
Protons per bunch                                  & $10^{11}$    &      1.7 \\  
Transverse emittances                                  &  mm.mrad    &     20      \\  
$\beta^*$ at IP6/IP8                               &     m       &      0.65   \\  
Longitudinal emittances                                   &  eV.s       &     2      \\ 
Voltage of 28~MHz RF                                &  kV         &     360    \\  
Voltage of 197~MHz RF                                &  kV         &     300    \\  
R.m.s.\@ momentum spread                                &  $10^{-4}$   &     1.7    \\  
R.m.s.\@ bunch length                                   &   m         &    0.45    \\  
Beam--beam parameter per IP                         &    -         &   0.007   \\  
Hourglass factor                                  &    -         &   0.85    \\  
Peak luminosity                                    & $10^{30}$cm$^{-2}$s$^{-1}$  &   165  \\
\hline\hline
\end{tabular}
\end{table}

\subsection{Beam Lifetime}

In the previous RHIC p-p runs, after the beams were brought into collision, we normally observed a large beam loss in the first hour, followed by a slow beam loss in the rest of store.  At the beginning, the instant maximum beam loss rate could reach 30\% per hour.  The beam loss rate of the slow loss was  typically 1-2\% per hour.  The burn-off contribution to the beam loss rate is less than 1\% per hour.

Empirically, the total beam intensity can be fitted with double exponentials~\cite{Lifetime}:
\begin{equation}
N_p(t)=A_1\exp(-t/\tau_{1})+ A_2\exp(-t/\tau_{2}),
\end{equation}
where $N_p(t)$ is the bunch intensity, and $A_{1,2}$ and  $\tau_{1,2}$ are fit parameters.  Figure~3 shows an example of  beam intensity evolution at store that was fitted with Eq.~(3).  Figure~4 shows  ($\tau_1$,$\tau_2$) of the physics stores in the past three 250~GeV or 255~GeV p-p runs. The reasons for the fast and slow beam losses will be discussed in the next section.

\begin{figure} [!]
\begin{center}
\includegraphics[width=70mm]{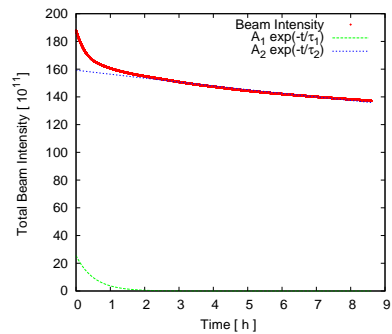}
\end{center}
\caption{An example of the Blue ring beam intensity evolution at store that was fitted with  Eq.~(3). The fill number is 16697.}
\end{figure}

\begin{figure} [!]
\begin{center}
\includegraphics[width=70mm]{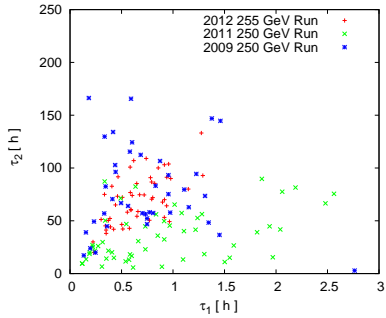}
\end{center}
\caption{The Blue ring beam loss fit parameters $\tau_1$ and $\tau_2$ in the previous 250~GeV runs.  }
\end{figure}

\subsection{Transverse Emittance and Bunch Length}

The transverse emittances are routinely measured with Ionization Profile Monitors (IPMs) in the RHIC. Figure~5 shows an example of the IPM-measured emittances at store. IPMs require knowledge of $\beta$ functions and need periodic calibrations of micro-channels.  An averaged all-plane emittance of both rings can be derived from luminosity based on Eq.~(1). In the previous RHIC p-p runs,  after beams were brought into collision, the measured emittances decreased in the first hour and then slowly increased in the rest of store.  The early emittance reduction was related to the large beam loss at the beginning of store.  Experiments showed that the emittance growth with beam--beam was smaller than that without beam--beam.

Bunch length was measured with a Wall Current Monitor (WCM). Figure~6 shows one example of averaged Full Width Half Maximum (FWHM) bunch lengths at store.  The spikes around 0.4 h in the plot were due to polarization measurement. To improve the signal-to-noise ratio during the polarization measurement, the voltages of the 197 MHz RF cavities were temporarily reduced to 20~kV. After the beams were brought into collision, the bunch length decreased in the first hour and then slowly increased in the rest of store.   The early bunch length reduction was related to the large beam loss at the beginning of store, and the bunch length growth with beam--beam was less than that without  beam--beam interaction.

\begin{figure} [!]
\begin{center}
\includegraphics[width=70mm]{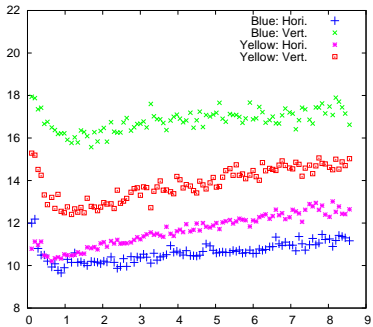}
\end{center}
\caption{IPM-measured transverse emittances at store for fill 16697.}
\end{figure}

\begin{figure} [!]
\begin{center}
\includegraphics[width=70mm]{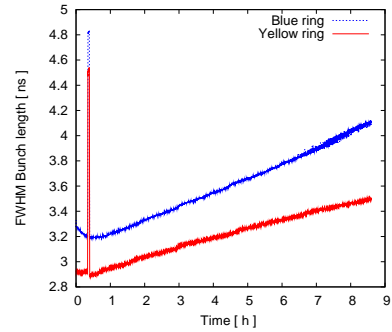}
\end{center}
\caption{WCM-measured FWHM bunch length at store for fill 16697.}
\end{figure}

\section{EXPLANATIONS}

\subsection{The Beam Loss Rate and Beam-Beam}

The store beam loss rate was mainly determined by the beam--beam interaction. Without beam--beam interaction, the beam loss rate can be better than  1\% per hour, depending on the machine tuning. After the beams were brought into collision,  the instant beam loss rate could reach a maximum of 30\% per hour.  The beam loss rates for bunches with one and two collisions were different. In the RHIC, 10 out of 109 bunches have only one collision instead of two collisions.  For example, for fill 15386, for bunches with one collision, the fitted $(\tau_1, \tau_2)$ are (1.5~h, 100~h), while for the bunches with two collisions, they are (0.8 h, 30 h).

\subsection{The Particle Loss Mechanism}

The WCM profile is actually the particle population distribution in the longitudinal plane. For a given period, we can calculate the number of particles leaking out of the central bunch area. Figure~7 shows each bunch's particle leakage percentage out of the [$-5$~ns, 5~ns] area during the first 0.5 h in the Blue ring in fill 15386, together with the actual particle loss percentage in the same period. The patterns of particle leakage and particle loss show a strong linear correlation, as shown in Fig. 8.

The strong linear correlation between particle leakage and particle loss is also true for the rest of store.  During the RHIC p-p runs, there was no de-bunched beam observed from the WCM profiles. Considering that particles in the bunch tail have larger off-momentum deviation, we conclude that the particles got lost in the transverse plane  due to a small transverse off-momentum dynamic aperture with beam--beam interaction.

\subsection{The Off-momentum Dynamic Aperture}

To achieve a short bunch length at physics store, 197~MHz RF cavities were used besides the acceleration RF cavities of 28~MHz. Figure~9 shows a typical longitudinal bunch profile.  With 197~MHz cavities, the relative momentum spread for the centre bucket between [$-2.5$~ns, 2.5~ns] increases to $5\times10^{-4}$. And for the tail particles out of [$-6$~ns, 6~ns] (full width), the relative momentum deviation is greater than $6\times10^{-4}$.

Figure~10 shows the calculated off-momentum dynamic aperture without and with beam--beam interaction from a $10^6$-turn particle tracking. The 2012 255~GeV Yellow ring lattice is used. The off-momentum dynamic aperture  with beam--beam is much smaller than that without beam--beam when the relative off-momentum deviation  $\mathrm{d}p/p_0>4\times10^{-4}$.  For the tail particles with  $\mathrm{d}p/p_0>6\times10^{-4}$, the dynamic aperture is less than 5~$\sigma$.

The large beam loss at the beginning of store was related to the initial large number of particles with large momentum deviation. Those particles were generated during RF re-bucketing and 197 MHz RF cavity voltage ramp-up. From WCM profiles, large-momentum particles  were observed on both sides of the centre bunch area after re-bucketing. We also observed beam loss shortly after RF re-bucketing without beam--beam.  When beams were brought into collision, the transverse off-momentum aperture was reduced. Those particles would get lost sooner or later in the first hour, depending on how close their $(\mathrm{d}p/p_0)_{max}$  to the off-momentum aperture and the longitudinal diffusion rate. 

\begin{figure} [!]
\begin{center}
\includegraphics[width=60mm]{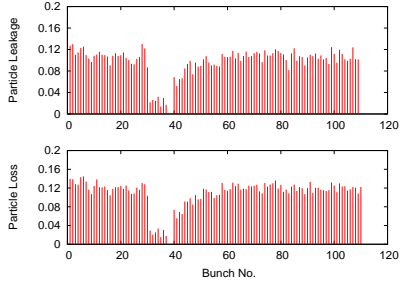}
\end{center}
\caption{The particle leakage and particle loss of all bunches in fill 15386.}
\end{figure}

\begin{figure} [!]
\begin{center}
\includegraphics[width=70mm]{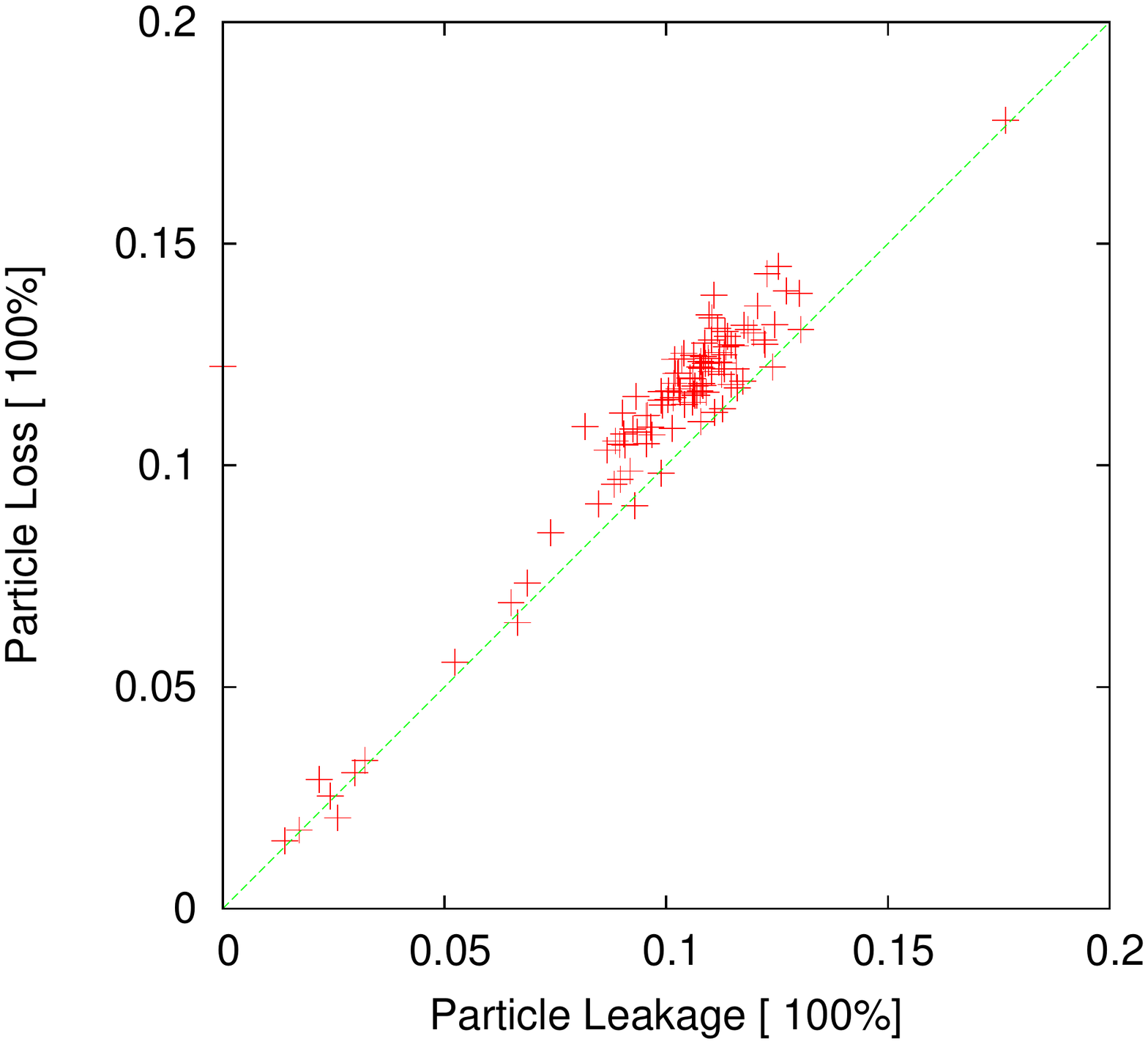}
\end{center}
\caption{Correlation of the particle leakage and particle loss of all bunches in fill 15386.}
\end{figure}

\begin{figure} [!]
\begin{center}
\includegraphics[width=70mm]{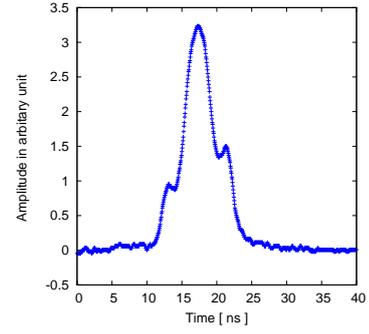}
\end{center}
\caption{Longitudinal bunch profiles with RF re-bucketing and 197 MHz RF voltages.}
\end{figure}

\begin{figure} [!]
\begin{center}
\includegraphics[width=70mm]{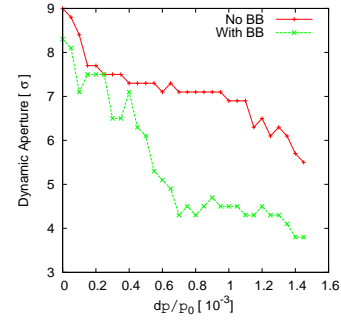}
\end{center}
\caption{Calculated off-momentum apertures without and with beam--beam interaction.}
\end{figure}

\subsection{Intra-beam Scattering Effects}

The slow loss after the first hour into store was linked to slow diffusion processes.  Here, we calculate the effects of  Intra-Beam Scattering (IBS) on the proton beam emittance and bunch length growth. With a smooth ring approximation,  the longitudinal and transverse IBS growth rates can be calculated as follows~\cite{IBS-Alexei}:
\begin{equation}
\tau^{-1}_{||} =\frac{1}{\sigma^2_p} \frac{\mathrm{d}\sigma^2_p}{\mathrm{d}t} \frac{r^2_i c N_i \Lambda}{8\beta \gamma^3 \epsilon^{3/2}_x<\beta_x^{1/2}> \sqrt{\pi/2}\sigma_l\sigma^2_p},
\end{equation}
\begin{equation}
\tau_{\bot} = \frac{\sigma^2_p}{\epsilon_x} <\frac{H_x}{\beta_x} > \tau^{-1}_{||}.
\end{equation}
Here, $\sigma_l$ and $\sigma_p$ are the r.m.s.\@ bunch length and the r.m.s.\@ relative  momentum spread. $H_x = \gamma_x D_x^2 + 2 \alpha_x D_x D'_x + \beta_x D'^2_x$, where $\alpha_x, \beta_x, and \gamma_x$ are Twiss parameters. $D_x$ and $D'_x$ are the horizontal dispersion and its derivative. $\Lambda$ is the Coulomb logarithm.

Based on Eqs.~(2) and (3), Figs.~11 and ~12 show an example  of the IBS contributions to the emittance and bunch length for fill 16697. We took the bunch intensity evolution, the initial emittance, and the bunch length from the real measurements. Comparing the calculated IBS contributions  to the
luminosity-derived emittance and the WCM-measured bunch length, the emittance and bunch length growth after 1.5 h into store are largely consistent with IBS.

\begin{figure} [!]
\begin{center}
\includegraphics[width=70mm]{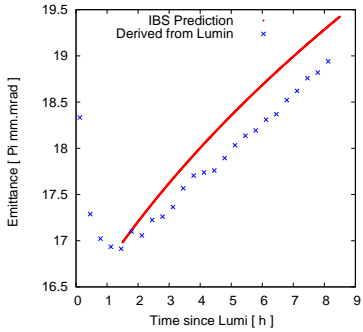}
\end{center}
\caption{Emittance modelling with IBS for fill 16697, compared to the emittance derived from luminosity.}
\end{figure}

\begin{figure} [!]
\begin{center}
\includegraphics[width=70mm]{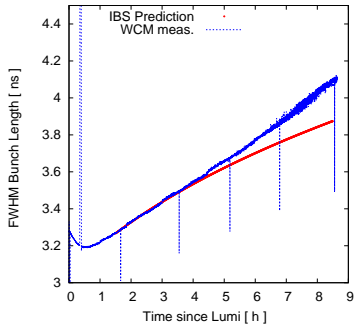}
\end{center}
\caption{Bunch length modelling with IBS for fill 16697, compared to the WCM bunch length measurement.}
\end{figure}

\section{LIMITS}

\subsection{Low-$\beta^*$ Lattices}

In order to further increase the luminosity, we can either increase the bunch intensity or reduce $\beta^*$.  A low-$\beta^*$ lattice increases the $\beta$ functions in the triplet quadrupoles, and therefore the particles will sample large non-linear magnetic field  errors at these locations. As a result, the dynamic aperture will be reduced~\cite{DA100-1}. For example, in the 2009 100~GeV p-p run, we used a lattice with $\beta^*=0.7$~m, which gave a short beam lifetime~\cite{RHIC2009}. At 250~GeV, we achieved $\beta^*=0.65$~m. The reason is that the transverse beam size is smaller at 250~GeV than at 100~GeV~\cite{DA100-2}.

A low-$\beta^*$ lattice also increases the non-linear chromaticity and reduces the off-momentum dynamic aperture. Chromatic analysis shows that the non-linear chromaticities are mostly originating from the
low-$\beta$ insertions IR6 and IR8~\cite{chrom1}. The non-linear chromaticities increase dramatically with the decreased $\beta^*$.   Figure~13 shows the calculated second- and third-order chromaticities as functions of $\beta^*$. Large second-order chromaticities push the particles with large momentum errors to the third- or 10th-order resonances. Several correction techniques for non-linear chromaticities have been tested and implemented in the RHIC~\cite{nonlinear}. To further reduce $\beta^*$, we need to balance the hourglass effect, beam lifetime reduction, and the luminosity increase. 

\begin{figure}[!]
\begin{center}
\includegraphics[width=70mm]{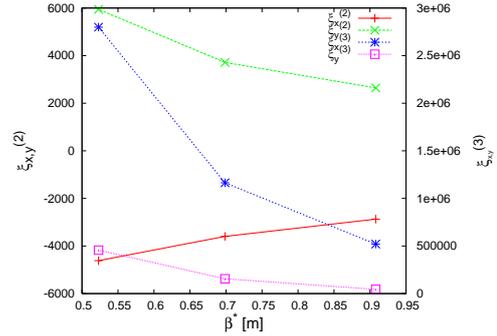}
\end{center}
\caption{Calculated second- and third-order chromaticities versus $\beta^*$ in the p-p runs.}
\end{figure}

\begin{figure}[!]
\begin{center}
\includegraphics[width=70mm]{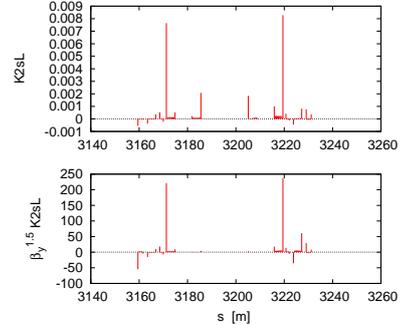}
\end{center}
\caption{$K_{2s}L$ and $\beta_y^{1.5}K_{2s}L$ in IR8 in the Yellow ring, based on the offline non-linear IR model.}
\end{figure}

\begin{figure}[!]
\begin{center}
\includegraphics[width=70mm]{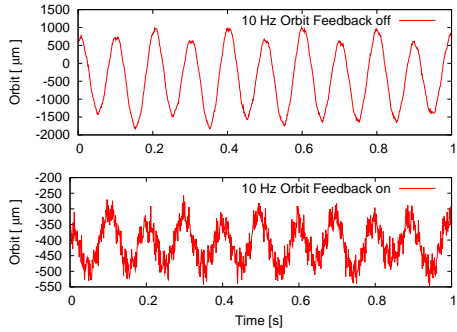}
\end{center}
\caption{Orbit oscillation without and with 10~Hz orbit feedback. The fill number is 15257.}
\end{figure}

\subsection{$3Q_{x,y}$ Resonances}

To mitigate the coupling between two beams, we would like to mirror the working points of the two RHIC rings on both sides of the diagonal in the tune space. However, in the operation we had to operate with both working points below the diagonal for better beam lifetimes. It was understood in the 2006 100~GeV p-p run that the strong $3Q_x$ resonances at $Q_x=2/3$ prevented a working point above the diagonal~\cite{RHIC2006}. At that time, the proton bunch intensity was $1.3\times10^{11}$.

In the 2012 100~GeV run, even with both working points below the diagonal, when the bunch intensity was higher than $1.7\times10^{11}$, we observed a larger beam loss due to the $3Q_y$ resonance, which is located at $Q_y=2/3$~\cite{RHIC2012}. The main contributions to the third-order resonances are from the sextupole and skew sextupole components in IR6 and IR8.  As an example, Fig.~14 shows $K_{2s}L$ and $\beta_y^{1.5}K_{2s}L$ in IR8 in the Yellow ring. To reduce the resonance stop-bands,  we routinely correct the  local sextupole and skew sextupole errors with IR orbit bumps by  minimizing the feed-down tune shifts~\cite{IRBUMP}, which improved the beam losses experimentally. Measurement and correction of the global third-order resonance driving terms with a.c. dipole excitation were also applied~\cite{RDT1, RDT2}.

\subsection{The 10~Hz Orbit Oscillation}

At the beginning of the 2008 100~GeV p-p run, we tested a near-integer working point (0.96, 0.95) in the Blue ring while keeping the working point in the Yellow ring at (0.695, 0.685). Weak--strong beam--beam simulation shows that there is a wider tune space with good dynamic apertures than the working point (0.695, 0.685)~\cite{DAinteger}. The spin simulation shows that there are weaker spin depolarization resonances in this region as well.

However, operating at near-integer tunes  turned out to be very challenging~\cite{RHIC2008}. With such tunes, we found that it was difficult to correct the closed orbit and to control the $\beta$-beat. Moreover, both detectors reported high background rates from the beam in the Blue ring when two beams were brought into collision. These backgrounds were caused by horizontal orbit vibrations around 10~Hz, which originated from mechanical vibrations of the low-$\beta$ triplets driven by the cryogenic flow~\cite{10Hz-1}.

We were able to correct the 10~Hz orbit oscillations in the 2011 p-p run by developing a local 10~Hz orbit feedback system~\cite{Robs}.  Figure 15 shows an example of horizontal Beam Position Monitor (BPM) readings in the triplet without and with the 10~Hz orbit feedback. The  peak-to-peak amplitude of the 10~Hz orbit oscillation was reduced by the feedback system from 2500~$\mu$m down to 250~$\mu$m.  We plan to revisit the near-integer working point in future beam experiment sessions.

\section{SUMMARY}

In this article, we have reviewed the beam--beam observations in the previous polarized proton runs in the RHIC. Particle loss happened in the transverse plane, and was due to the limited transverse off-momentum dynamic aperture. Beam--beam interaction, IR non-linear multipole field errors, non-linear chromaticities with low $\beta^*$s, and  $3Q_{x,y}$ resonances reduce the transverse dynamic aperture.  Measures had been implemented in the RHIC to correct the  non-linear chromaticities and  $3Q_{x,y}$ resonance driving terms.  The 10~Hz orbit modulation was reduced by means of a 10 Hz orbit feedback.  To further increase the luminosity in the RHIC p-p operations, we plan to increase the bunch intensity and to reduce $\beta^*$ at collisional IPs. To reduce the large beam--beam tune spread from high bunch intensities, head-on beam--beam compensation with electron lenses is being installed in the RHIC.

\section{ACKNOWLEDGEMENTS}

The authors would like to thank M. Bai, M. Blaskiewicz, H. Huang, M. Minty, C. Montag, V. Ptitsyn, V. Scheoffer, and S. White for stimulating discussions during this work.


\end{document}